\documentclass[prb,twocolumn,nopacs,floatfix,superscriptaddress]{revtex4-2}       
   
\usepackage{amssymb}
\usepackage{amsmath}       
\usepackage{graphicx}
\usepackage{color}
\usepackage{enumerate}
\usepackage[normalem]{ulem}
\usepackage[colorlinks,bookmarks=false,citecolor=blue,linkcolor=blue,urlcolor=blue]{hyperref}

\begin{document}

\linepenalty=10
\interlinepenalty=100
\clubpenalty=5000
\widowpenalty=5000
\brokenpenalty=1000
\tolerance=900
\hyphenpenalty=100000

\title{\boldmath Switching of antiferromagnetic states in LiCoPO$_4$ as investigated via the magnetoelectric effect\unboldmath}

\author{Vilmos Kocsis}
\altaffiliation[Current address: ]{Institut f\"ur Festk\"orperforschung, Leibniz IFW Dresden, 01069 Dresden, Germany}
\affiliation{RIKEN Center for Emergent Matter Science (CEMS), Wako, Saitama 351-0198, Japan}

\author{Yusuke Tokunaga}
\affiliation{RIKEN Center for Emergent Matter Science (CEMS), Wako, Saitama 351-0198, Japan}
\affiliation{Department of Advanced Materials Science, University of Tokyo, Kashiwa 277-8561, Japan}

\author{Yoshinori Tokura}
\affiliation{RIKEN Center for Emergent Matter Science (CEMS), Wako, Saitama 351-0198, Japan}
\affiliation{Tokyo College and Department of Applied Physics, University of Tokyo, Hongo, Tokyo 113-8656, Japan}

\author{Yasujiro Taguchi}
\affiliation{RIKEN Center for Emergent Matter Science (CEMS), Wako, Saitama 351-0198, Japan}


\begin{abstract}
The linear magnetoelectric (ME) effect allows for the selection or switching between two antiferromagnetic (AFM) states via the application of large electric ($E$) and magnetic ($H$) fields.
Once an AFM state is selected, it is preserved by an energy barrier, even when the fields are removed.
Using a simple phenomenological model, we find that this energy barrier, needed to switch the AFM state, is proportional to the product of the $E$ and $H$ coercive fields $(EH)_{\rm C}$.
We measured the field and temperature dependence of $(EH)_{\rm C}$ in LiCoPO$_4$ for two different field configurations, and the data show the temperature variation of $(EH)_{\rm C}\sim(T_{\rm N}-T)^{3/2}$ in good agreement with the model.
We also investigated the dynamics of the AFM domain switching using pulsed $E$-field measurements.
It was found that the coercive field $(EH)_{\rm C}$ follows a power-law frequency dependence and is well described in the framework of Ishibashi-Orihara model, implying 1-dimensional character of domain wall propagation.
\end{abstract}

\maketitle

%
%
	\section{Introduction} 

    \begin{figure}[t!]
 	
    \includegraphics[width=8.5truecm]{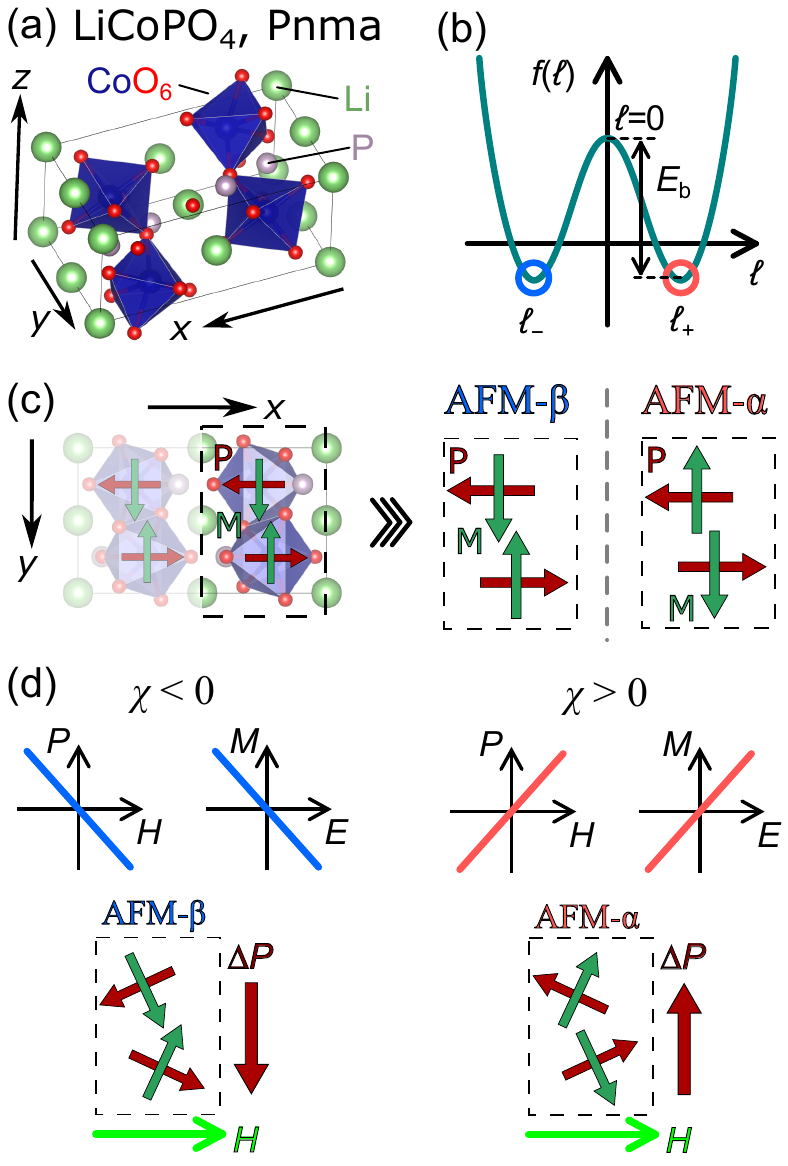}
    \caption{(Color online)
        (a) Schematic unit cell of LiCoPO$_4$. 
        (b) Schematic of the free energy landscape of a uniaxial antiferromagnet. The order parameter is the staggered magnetic moment ($\ell$) and the free energy has two stable ($\ell_\pm$) and one unstable ($\ell=0$) extrema.
    	The two possible states, AFM-$\alpha$ ($\ell_+$) and AFM-$\beta$ ($\ell_-$), are separated by an energy barrier $E_b$.
    	(c) The magnetic unit cell and the AFM states constituted by four Co$^{2+}$ ions with $S$=3/2 are illustrated by a representative segment of the magnetic unit cell. Green and red arrows are the local magnetization and polarization vectors, respectively.
    	(d) The two AFM domains are equivalent to the two ME-states with positive and negative ME-susceptibilities, respectively.
    	The local magnetic moments cant towards the applied $H$-field, while the local $P$ rotate in opposite directions.
}
    \label{lcpo01a}
    \end{figure}

Electric ($E$) and magnetic field ($H$) control over the antiferromagnetic (AFM) order parameter is one of the most intriguing aspects of spintronic applications~\cite{Fiebig2002a,Ramesh2007,Yamasaki2007,Taniguchi2008,Seki2009PRL,Garcia1106,Zhou2010,Tokura2010,Tokura2014,Matsukura2015NatNano,Jungwirth2016,Shiratsuchi2018APL,Baltz2018RMP,Kocsis2018PRL}.
Contrary to ferromagnetic devices, an antiferromagnet based architecture is protected against stray $H$ fields, and offers a faster operation with less power consumption~\cite{Bibes2008,Chu2008,Heron2014a,Heron2014,Fusil2014}.
In itinerant antiferromagnets, the order parameter is manipulated using spin-polarized currents via the spin-transfer torque mechanism~\cite{MacDonald2011}.
In case of AFM insulators, manipulation of the magnetic order can be achieved via the magnetoelectric (ME) effect without using electric current and hence without generating joule heating.
In ME materials, electric polarization ($P$) and magnetization ($M$) are induced by the application of $H$ and $E$ fields, respectively~\cite{Katsura2005,Jia2006,Jia2007}.
Using the combination of high $E$ and $H$ fields, it is possible to form a single-domain AFM state with its $P$ and $M$ parallel to the fields, respectively~\cite{Kocsis2018PRL}.
From the viewpoint of applications, particularly in memory devices, the stability of this AFM state plays a key role.
The magnetic state has to be robust against external stimuli as well as thermal agitation, while it is advantageous if the AFM state can be selected and/or switched by simultaneous application of small $E$ and $H$ fields.
Therefore, we investigated the stability and switching characteristics of the AFM state for a prototypical ME material, LiCoPO$_4$.

    \begin{figure}[t!]
 	
    \includegraphics[width=8.5truecm]{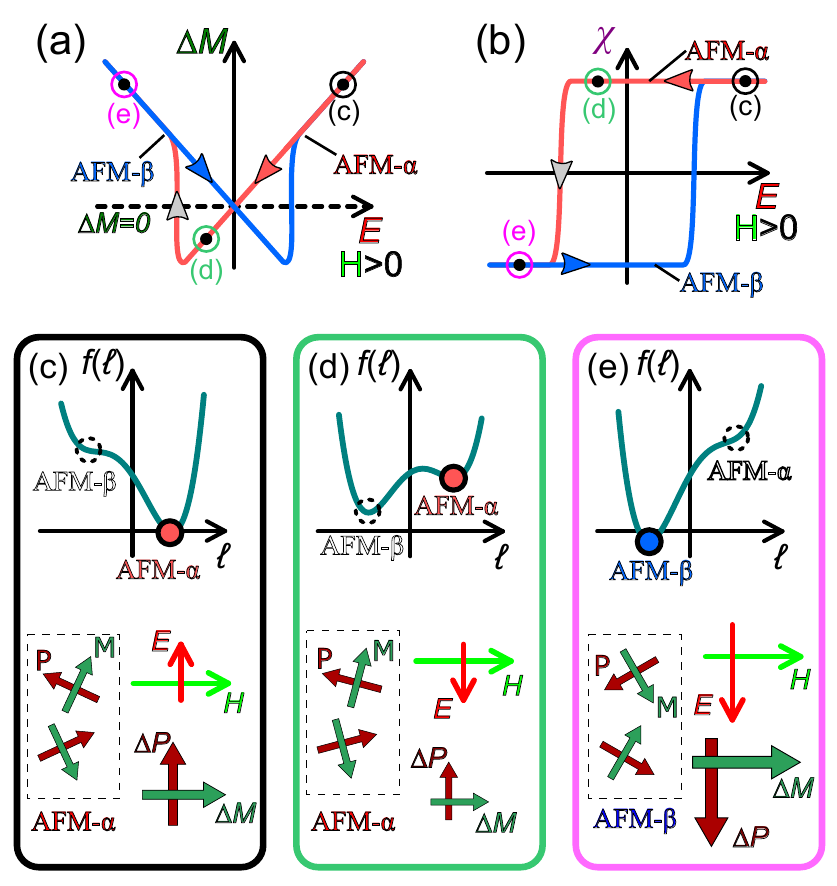}
    \caption{(Color online)
		(a) $M$-$E$ loop in the presence of $H$-field, where the magnetization change $\Delta{M}$ is measured from the $E$=0 state. In the absence of $H$ field the $M$-$E$ loop is linear, while in the presence of the $H$-field the $M$-$E$ loop has symmetric, butterfly shape due to the switching of the ME states between $\alpha$ and $\beta$.
		(b) Correspondingly, the ME susceptibility in the presence of $E$ and $H$ fields shows the switching between positive (AFM-$\alpha$) and negative (AFM-$\beta$) values.
		(c-e) Schematic illustration of the switching process between the two ME states. (c) In the initial state the $E$ and $H$ fields select the AFM-$\alpha$ domain ($\mathbf{E}\parallel\mathbf{P}$, $\mathbf{H}\parallel\mathbf{M}$) with positive ME susceptibility. In this case the fields are high and there is no energy barrier.
		(d) When $E$ is reversed, the  free energy for AFM-$\alpha$ state is increased with the $+P\cdot{E}$ contribution, but it is protected by the energy barrier and preserved as a meta-stable state.
		(e) When the $E$ field is further increased with negative sign and becomes large enough, the energy barrier vanishes between the AFM-$\alpha$ and AFM-$\beta$ states, and the later magnetic state is stabilized.
}
    \label{lcpo01b}
    \end{figure}

LiCoPO$_4$ has a centrosymmetric and orthorhombic olivine-type structure (space group $Pnma$), as shown in Fig.~\ref{lcpo01a}(a).
Symmetry of the lattice allows a staggered order of local electric polarization lying within the $xz$ plane at the sites of four Co$^{2+}$ ions with one mirror plane.
Below $T_{\rm N}$=21.3\,K, the spins with $S$=3/2 order into a four-sublattice N\'eel-type AFM arrangement with the moments pointing along the $y$ axis~\cite{Santoro1966,Vaknin2002}.
The magnetic order simultaneously breaks the inversion and the time-reversal symmetries, allowing linear ME effect, $P_\mu=\chi_{\mu\nu}H_\nu$ and $\mu_0M_\mu=\chi_{\nu\mu}E_\nu$, where $\chi_{\nu\mu}$ is ME tensor components with $\mu,\nu = x,y$~\cite{Rivera1994}.
The ME properties of LiCoPO$_4$ have often been discussed in terms of the toroidal moment~\cite{Ederer2007,Aken2007,Zimmermann2014}, and previous laser-optical second-harmonic generation measurements have demonstrated the emergence of the ferrotoroidal order~\cite{Aken2007,Zimmermann2014}, while recent THz absorption measurements have found that LiCoPO$_4$ has also symmetric component of ME tensor, namely quadrupolar moment~\cite{Kocsis2019PRB}.

In a ME compound, switching of the ME state corresponds to the reversal of the toroidal and the quadrupolar moment of the unit cell.
In LiCoPO$_4$, this is realized via the reversal of the AFM state~\cite{Zimmermann2014,Kocsis2018PRL}.
Therefore, we discuss the change in the ME state of LiCoPO$_4$ in terms of the changes in the AFM state throughout the paper.
As shown in Figs.~\ref{lcpo01a}(b-d), the AFM order of LiCoPO$_4$ has two possible domains, AFM-$\alpha$ and AFM-$\beta$, which are also the two ME states with opposite sign of the ME tensor, $\chi_{\mu\nu}(\alpha) = -\chi_{\mu\nu}(\beta)$.
Note that besides the time reversal symmetry, the inversion symmetry also reverses the sign of the AFM order parameter as it exchanges Co ions with anti-parallel spins.
Single-domain AFM (and ME) state can be selected by cooling the crystals across $T_{\rm N}$ in the presence of both $E$ and $H$ fields in a crossed geometry ($\mathbf{E}\perp\mathbf{H}$, $\mathbf{E}$,$\mathbf{H}\perp{z}$)~\cite{Zimmermann2014,Kocsis2018PRL,Kocsis2019PRB}, which is so-called ME-poling.
Further details of the ME poling are discussed in relation to Fig.~S1~\cite{Kocsis2020PRB2SM}.

In this paper, we discuss the stability and switching characteristics of the two AFM states in LiCoPO$_4$ in terms of a simple phenomenological model, and both quasi-static and dynamic measurements of ME properties.
The phenomenological model of a ME antiferromagnet in both $E$ and $H$ fields gives a formula of the coercive product field in terms of energy barrier separating the two AFM states, ME susceptibility, and its temperature dependence.
Isothermal switching behavior of the AFM state similar to the previous optical imaging under static fields~\cite{Zimmermann2014} is observed through quasi-static ME measurements.
In addition, we present the temperature evolution of the coercive product field, which is found to be in good accord with the phenomenological model.
Switching dynamics of the AFM state investigated by using pulsed $E$-field measurements reveals one-dimensional feature of domain wall propagation.

    \begin{figure}[t!]
 	
    \includegraphics[width=8.0truecm]{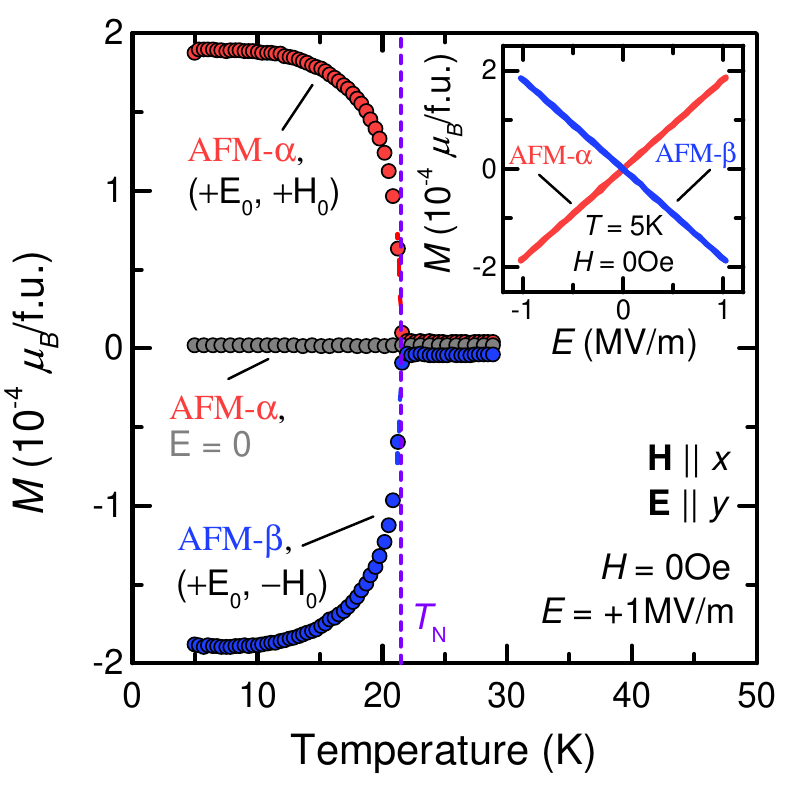}
    \caption{(Color online)
    	Temperature dependence of magnetization in the antiferromagnetic LiCoPO$_4$, measured in the absence of $H$ field for the warming runs.
    	The AFM-$\alpha$ state was stabilized by cooling the sample in the presence of ($+E_0$, $+H_0$) poling fields, while the AFM-$\beta$ state was prepared by using ($+E_0$, $-H_0$) fields, with $\vert{E}_0\vert$=1\,MV/m, $\vert{H}_0\vert$=30\,kOe, and $\mathbf{E}_0\parallel{y}$, $\mathbf{H}_0\parallel{x}$. In the presence of  $E>0$ field, the AFM-$\alpha$ domain has positive magnetization (red curve), while the AFM-$\beta$ state has negative magnetization (blue curve). In the absence of any field the magnetization is zero (grey curve).
    	The inset shows the $E$ field dependence of the magnetization of the AFM-$\alpha$ and AFM-$\beta$ states in the absence of $H$ field at $T$=5\,K.
    	}
    \label{lcpo02}
    \end{figure}

    \begin{figure}[t!]
 	
    \includegraphics[width=8.0truecm]{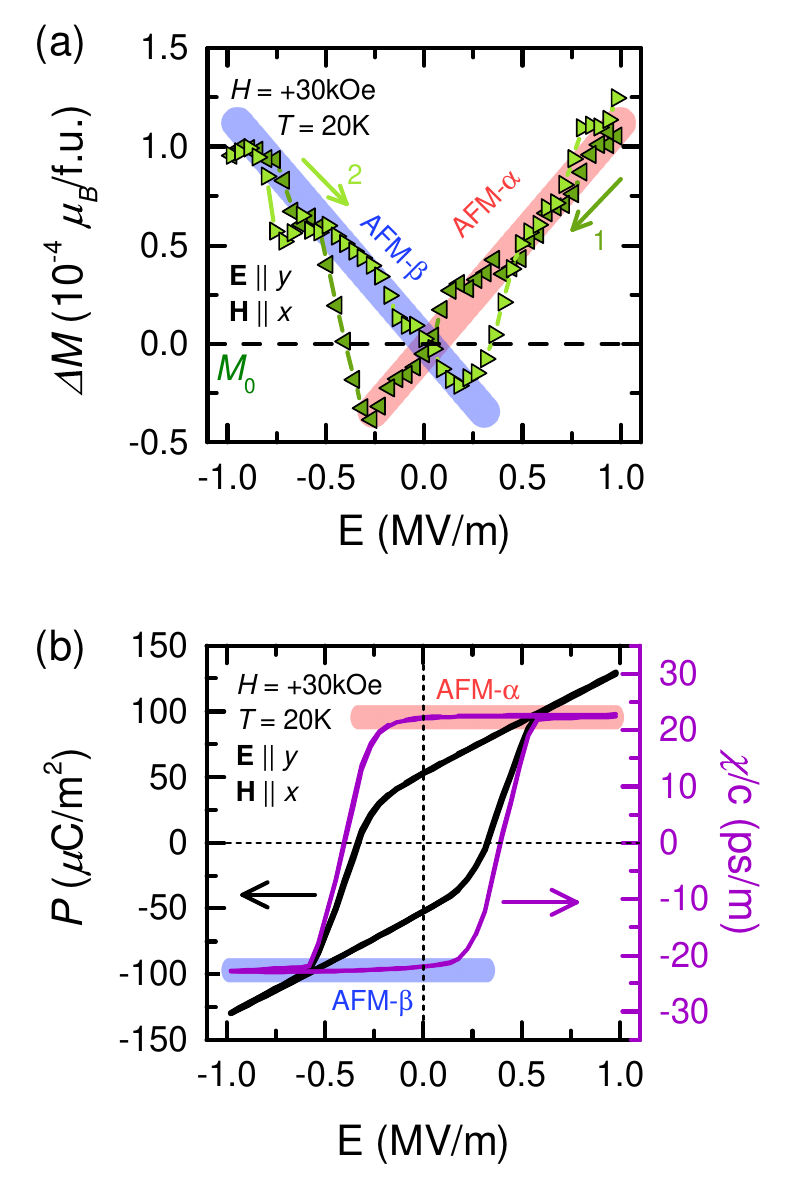}
    \caption{(Color online) (a) Electric field dependence of the magnetization, measured in the presence of $H$=30\,kOe field at $T$=20\,K. The fields were applied in the $\mathbf{E}\parallel{y}$ and $\mathbf{H}\parallel{x}$ configuration. The magnetization change ($\Delta{M}=M(E)-M_0$) is shown with respect to the base magnetization $M_0$ measured at $H$=+30\,kOe and $E$=0\,MV/m. (b) $P$-$E$ hysteresis loop of LiCoPO$_4$, measured in $H$=30\,kOe and $T$=20\,K. The magnetization and polarization measurements were done simultaneously. In both panels, ME susceptibility of the AFM-$\alpha$ and AFM-$\beta$ states are highlighted by red and blue markings, respectively. The butterfly-shaped $M$-$E$ loop is the result of the switching between the AFM-$\alpha$ and AFM-$\beta$ states.}
    \label{lcpo03}
    \end{figure}

%
%
	\section{Experimental Methods and the ME poling}
	
Single-crystalline LiCoPO$_4$ was grown by using the optical floating zone technique~\cite{SaintMartin2008}.
For the measurements of $P$ and $E$-field induced $M$, the single crystals were cut into 190\,$\mu$m thick slabs with $xz$ and $yz$ faces, and Au/Pt electrodes were sputtered.
The $E$ and $H$ fields were applied perpendicular to each other and to the $z$ axis, \textit{i.e.}, $\mathbf{E}\perp\mathbf{H}$, and $\mathbf{E}$, $\mathbf{H}\perp{z}$.
For the slabs with $xz$ faces, $\mathbf{E}\parallel{y}$ and $\mathbf{H}\parallel{x}$, while for those with $yz$ faces, $\mathbf{E}\parallel{x}$ and $\mathbf{H}\parallel{y}$ fields were applied.
For the $E$-field induced $M$ measurements, the whole surface of the slabs were polished, Au/Pt were sputtered, and then the edges were cut off into rectangular shape.

$H$-field dependent $P$ was measured in a Physical Property Measurement System (Quantum Design, PPMS) by integrating the displacement current with a capacitor (Q-mode) using an electrometer (Keithely, 6517A).
The $E$-field-dependent $M$ was measured using the same electrometer as a voltage source, while at the same time the $P$-$E$ hysteresis loops were measured.
The $H$ field and sample environment for these experiments was provided by a Magnetic Property Measurement System (Quantum Design, MPMS-XL).
The ferroelectric $P$ of magnetic origin was also measured with the so-called PUND (Positive-Up-Negative-Down) technique using pulsed $E$-fields. For these measurements we used a ferroelectric tester equipped with 500\,V amplifier (Radiant Inc., Precision Premier II.).

During the procedure of ME-poling, large $E$ and $H$ fields are applied when the samples are cooled across the magnetic ordering temperature.
The single ME domain state is prepared by the cross product of the poling fields, $\mathbf{E}_0$ and $\mathbf{H}_0$, applied parallel to polarization and magnetization, respectively.
There are four different combinations of the poling $E_0$ and $H_0$ fields to be considered with a fixed geometry of $\mathbf{E}_0\parallel{x}$, $\mathbf{H}_0\parallel{y}$.
In order to simplify the description of the experiments, we will refer to the $\mathbf{E}_0\parallel{+x}$, $\mathbf{H}_0\parallel{+y}$ combination of the poling fields as ($+E_0$, $+H_0$), while the $\mathbf{E}_0\parallel{-x}$, $\mathbf{H}_0\parallel{+y}$ combination is denoted as ($-E_0$, $+H_0$).
For the other geometry with 90$^\circ$ rotated fields, ($+E_0$, $+H_0$) indicates $\mathbf{E}_0\parallel{+y}$, $\mathbf{H}_0\parallel{-x}$, and ($-E_0$, $+H_0$) represents $\mathbf{E}_0\parallel{-y}$, $\mathbf{H}_0\parallel{-x}$.
Note, that throughout this paper the electric and magnetic coercive fields ($E_{\rm C}$ and $H_{\rm C}$) refer to the switching of the AFM state.

%
%
	\section{Phenomenological model of a ME antiferromagnet} \label{Sec:Theory}

In this section we describe a simplified phenomenological model, which illustrates the stability of an AFM order parameter in external $E$ and $H$ fields.
For simplicity, we take free energy expansion of a uni-axial AFM in transverse $H$ field.
The interaction between the order parameter and the fields is described by the quadratic expansion of the free energy, which accounts for the linear ME effect.
Although the model does not describe the particular case of LiCoPO$_4$ in an exact manner, it can give a clear insight and a reasonable interpretation for the experimental observations for the transverse field measurements ($\mathbf{H}\parallel{x}$).

The free energy of a uni-axial AFM is described by the following expansion in terms of the staggered moment ($\ell$) as an order parameter:
\begin{equation}
f_{AFM}(\ell) = f_0 -a\ell^2+b\ell^4,
\end{equation}
where $f_0$ is a constant contribution to the free energy, the parameter $a$ is assumed to have temperature dependence, $a(T)=a_0(T_{\rm N}-T)$, the $b$ is a constant, and both parameters are positive, $a_0>0$, $b>0$.
As illustrated in Fig.~\ref{lcpo01a}(b), the free energy has three extrema; One unstable at $\ell=0$, and two stable at $\ell_{\alpha / \beta}=\pm\sqrt{\frac{a}{2b}}$, corresponding to the AFM-$\alpha$ and AFM-$\beta$ states, respectively.
The two AFM states are separated by an energy barrier:
\begin{equation}
E_b = f_{AFM}(\ell=0) - f_{AFM}(\ell_{\alpha / \beta})= \frac{1}{4}\frac{a^2}{b},
\end{equation}
which vanishes at $T_{\rm N}$, following the temperature dependence of parameter $a$.
When one of the AFM states is selected, this energy barrier prevents it from changing to the other AFM state at enough low temperatures.

Interaction between the order parameter and the external fields $E$ and $H$ fields is described by the following second order expansion~\cite{Landau8,Noether1918,Spaldin2008,Mostovoy2010PRL,Mufti2011PRB}:
\begin{equation}
f_{EH}(\ell,E,H) = -\frac{1}{2}\mu H^2 -\frac{1}{2}\epsilon E^2 -g\ell{EH},
\end{equation}
where the parameter $g$ describes the ME coupling.
The last bilinear term is finite in LiCoPO$_4$, because both the $EH$ product and the order parameter $\ell$ changes sign for the time reversal and spatial inversion operations, leaving their product as an invariant.
In this model the ME susceptibility $\chi$ is proportional to the staggered moment $\ell$ with the coupling constant $g$, \textit{i.e.} $\chi = g \ell$:
\begin{equation}
P = -\frac{\partial{f}}{\partial{E}} = \epsilon E + g\ell H, \\ \label{eq:ME1}
\end{equation}
\begin{equation}
M = -\frac{\partial{f}}{\partial{H}} = \mu H + g\ell E.\label{eq:ME2}
\end{equation}
The resulting ME effect is linear in the fields, and the ME susceptibilities of the two AFM states have opposite signs ($\chi(\ell_{\alpha})=-\chi(\ell_{\beta}$)), as shown in Fig.~\ref{lcpo01a}(d).
Therefore, this model qualitatively describes the most important experimental characteristic of LiCoPO$_4$, namely, the presence of two AFM states with linear ME effect of opposing signs.

Simultaneous application of $E$ and $H$ fields deforms the free energy landscape, reduces the energy barrier $E_b$, and eventually switches the AFM states, as shown in Fig.~\ref{lcpo01b}.
The existence of the energy barrier depends on the number of extrema in the free energy landscape.
To find the critical $E$ and $H$ fields, where the energy barrier vanishes with an assumption of negligible thermal fluctuation, we consider the number of roots of the partial derivative:
\begin{equation}
\frac{\partial{f}}{\partial{\ell}} = -2a\ell +4b\ell^3-gEH,
\end{equation}
from which we take the cubic discriminant:
\begin{equation}
\Delta_f = 16b (8a^3-27bg^2H^2E^2).\label{eq:discr}
\end{equation}
For the application of low (or no) $E$ and $H$ fields, the discriminant is positive $\Delta_f > 0$, and the free energy landscape has two stable local minima and a finite energy barrier.
For the application of high $E$ and $H$ fields, the discriminant changes sign ($\Delta_f \leq 0$), the free energy has one minimum, and the energy barrier $E_b$ vanishes.
The condition of $\Delta_f = 0$ indicates the point at which one of the local minima vanishes, and should define the coercive $E$ and $H$ fields.
The product of the $E$ and $H$ coercive fields is expressed as:
\begin{equation}
(EH)_{\rm C} = \sqrt{\frac{8a^3}{27bg^2}}.\label{eq:EH1}
\end{equation}
This product coercive field can be also expressed in terms of the ME susceptibility and the zero-field energy barrier:
\begin{equation}
(EH)_{\rm C} = \frac{8}{\sqrt{27}} \frac{E_b}{\chi}.\label{eq:EH2}
\end{equation}
This relation makes a clear connection between the stability of the AFM states and the ME effect in the presence of external fields.
As expected, the product coercive field increases for larger energy barrier and for a weaker ME coupling.
In a good ME memory, the energy barrier should be large so that the selected AFM state is robust against either $E$ or $H$-field, as well as thermal agitation.
In addition, if the ME susceptibility is also large, then the strong coupling may significantly decrease the $(EH)_{\rm C}$ needed to switch the AFM state.
All in all, the $(EH)_{\rm C}$ product field is proportional to the ratio of the $E_b$ and $\chi$, and both of them should be preferably large for the applications.
Near $T\approx{T_{\rm N}}$, we put $a(T)=a_0(T_{\rm N}-T)$ into Eq.~\ref{eq:EH1}, and obtain:
\begin{equation}
(EH)_{\rm C} = \sqrt{\frac{8a_0^3}{27bg^2}}\left(T_{\rm N}-T\right)^\frac{3}{2}.\label{eq:EH3}
\end{equation}

    \begin{figure*}[t!]
 	
    \includegraphics[width=16.0truecm]{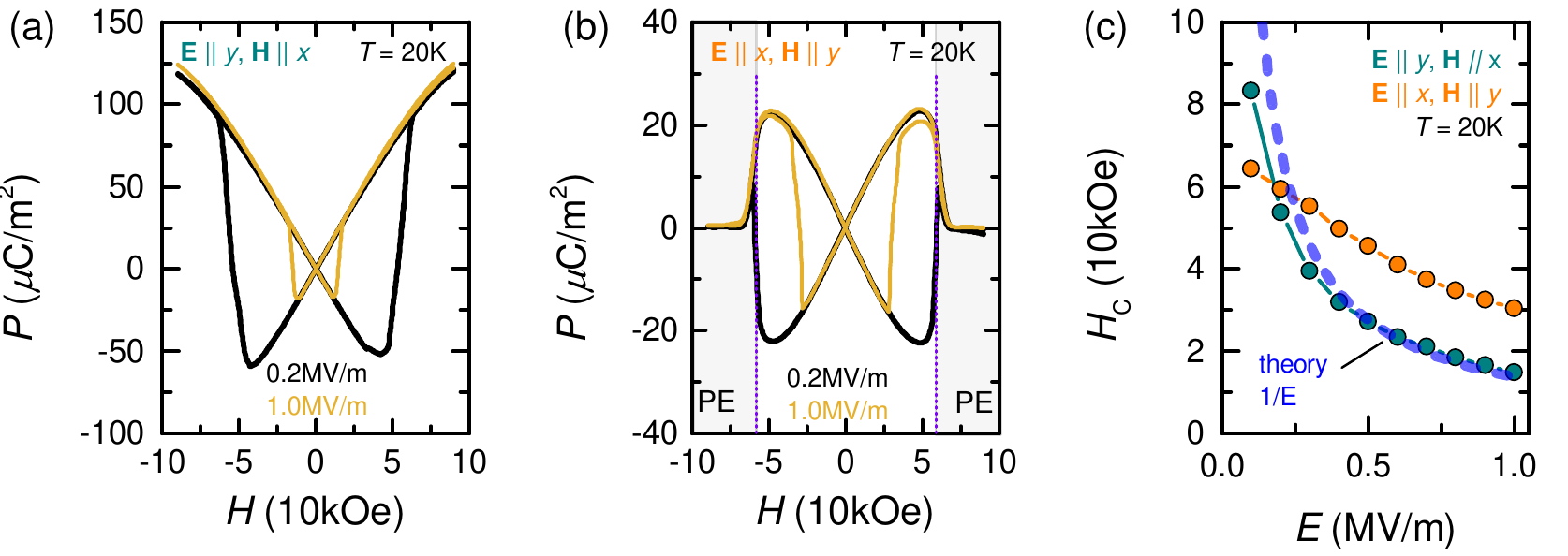}
    \caption{(Color online)
    (a,b) $P$-$H$ loops measured in the presence of $E$ field at $T$=20\,K. Measurement configurations are $\mathbf{E}\parallel{y}$, $\mathbf{H}\parallel{x}$, and $\mathbf{E}\parallel{x}$, $\mathbf{H}\parallel{y}$, respectively. Similarly to the $M$-$E$ hysteresis, the butterfly-shaped $P$-$H$ loops indicate the switching of the AFM states. In panel (b), the phase boundary of the $H$-field induced paraelectric (PE) state is indicated by grey shading and dashed line.
    (c) Bias $E$-field dependence of the coercive fields $H_{\rm C}$ for $\mathbf{E}\parallel{y}$, $\mathbf{H}\parallel{x}$, and $\mathbf{E}\parallel{x}$, $\mathbf{H}\parallel{y}$ configurations. The experimental data for the transverse field case, $\mathbf{E}\parallel{y}$, $\mathbf{H}\parallel{x}$, are well reproduced by the 1/$E$ dependence, as predicted by the theoretical model (dashed blue line).
    }
    \label{lcpo04}
    \end{figure*}

    \begin{figure*}[t!]
 	
    \includegraphics[width=16.0truecm]{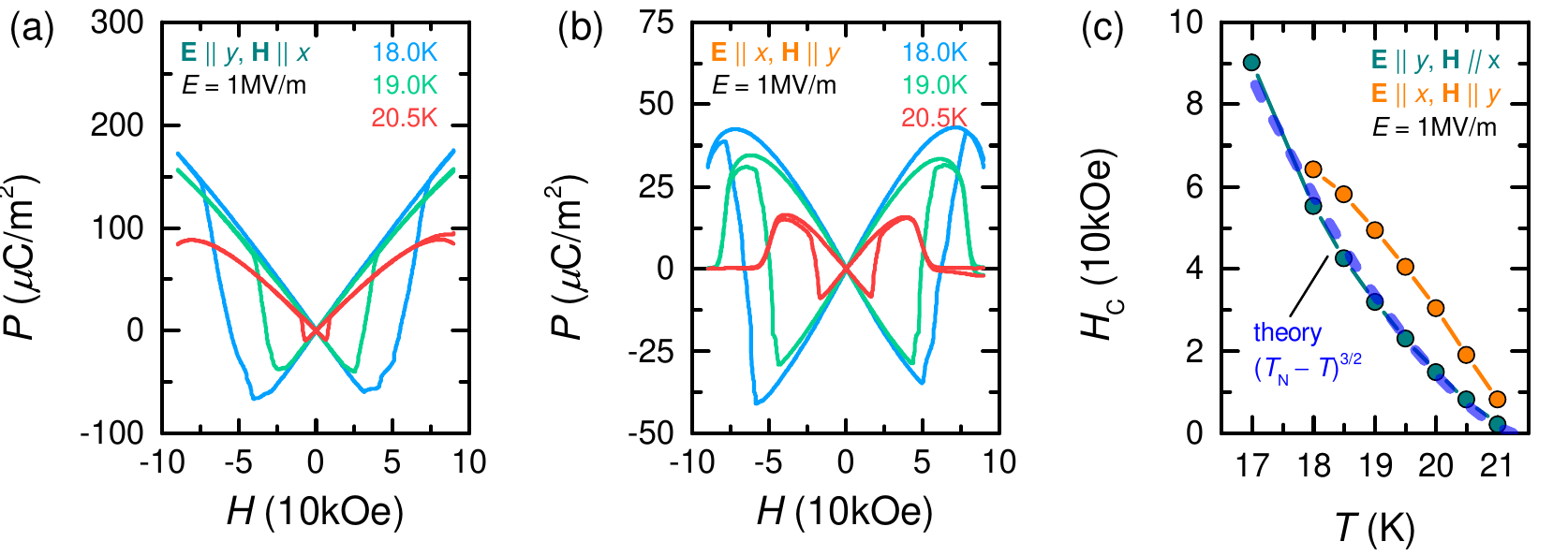}
    \caption{(Color online)
    (a,b) $P$-$H$ loops measured in the presence of $E$=+1\,MV/m field at selected temperatures.
    In panels (a) and (b), measurement geometries are $\mathbf{E}\parallel{y}$, $\mathbf{H}\parallel{x}$, and $\mathbf{E}\parallel{x}$, $\mathbf{H}\parallel{y}$, respectively.
    (c) Temperature dependence of the coercive field $H_{\rm C}$ in $E$=+1\,MV/m field. The theoretical model (dashed blue line) for the transverse field case, $\mathbf{E}\parallel{y}$, $\mathbf{H}\parallel{x}$, shows good agreement with the experimental data.}
    \label{lcpo05}
    \end{figure*}

%
%
	\section{Magnetoelectric effect and control over the AFM state}
 
A less investigated feature of a ME antiferromagnet is the $E$-field-induced $M$, \textit{i.e.} the converse ME effect.
Figure~\ref{lcpo02} shows the temperature dependence of the $M$ measured in the absence of $H$ field, but in the presence of large $E$ field.
The single domain ME-state (AFM-$\alpha$ or AFM-$\beta$) was initially selected by cooling the LiCoPO$_4$ single crystal in the presence of ($+E_0$, $+H_0$) or ($-E_0$, $+H_0$) field, \textit{i.e.} with $\mathbf{E}\parallel{y}$ and $\mathbf{H}\parallel{x}$.
The measurements were conducted in the warming runs after the removal of the $H$ field.
In the absence of both fields ($E=0$ and $H=0$, grey curve), the single domain state shows zero magnetization.
In the presence of high field $E$=+1\,MV/m, LiCoPO$_4$ has a weak magnetization, which disappears at $T_{\rm N}$.
On the basis of the measurement of the $E$-field induced $M$ (the inset to Fig.~\ref{lcpo02}) magnitude of the converse ME effect is evaluated to be $\chi_{xy}/c$=33.1\,ps/m ($c$ is the speed of light), in agreement with former reports of the $P$-$H$ measurements~\cite{Rivera1994,Kocsis2018PRL}.
Sign of the ME effect depends on the selected ME state; The AFM-$\alpha$ state selected by ($+E_0$, $+H_0$) has positive magnetization in $E$=+1\,MV/m field, and positive ME susceptibility.
When the other ME state (AFM-$\beta$) is selected by the ($-E_0$, $+H_0$) poling fields, sign of the ME effect is reversed to negative.
Note that the weak magnetization is linear in the $E$-field as $\mu_0M_x = \chi_{xy}E_y$, while sign of the ME susceptibility is governed by the AFM state ($\chi = g \ell$, $\ell_{\alpha}=-\ell_{\beta}$).

\subsection*{Stability of AFM state in static fields}

Antiferromagnetic states are inherently robust against the application of either $E$ or $H$ field, as the conjugate field to the AFM order parameter is the staggered magnetic field.
We could not observe the switching between the AFM states by applying the highest fields possible, either $E$=+2\,MV/m or $H$=+140\,kOe.
However, as discussed in Sec.~\ref{Sec:Theory}, the product of the external fields  $EH$ works as a conjugate field to the AFM order parameter via the ME effect.
The AFM states in LiCoPO$_4$ are manipulated by the simultaneous application of the $E$ and $H$ fields in a $\mathbf{E}\parallel{y}$, $\mathbf{H}\parallel{x}$ or $\mathbf{E}\parallel{x}$, $\mathbf{H}\parallel{y}$ geometry, as shown below.

In Fig.~\ref{lcpo03}(a) we show the $E$-field induced change in the magnetization $\Delta{M}=M(E)-M_0$, where $M_0$=0.13\,$\mu_B$/f.u. is the base magnetization measured at $H$=+30\,kOe and $E$=0\,MV/m.
For the application of large $E$ and $H$ fields at temperatures near $T_{\rm N}$, linear behavior of the ME effect is lost and the $\Delta{M}$-$E$ loops develop butterfly shape.
The experiment was started from a single-domain state, prepared by cooling the sample to $T$=20\,K in the presence of $E_0$=+1\,MV/m and $H_0$=+30\,kOe fields, $\mathbf{E}\parallel{y}$, $\mathbf{H}\parallel{x}$.
During the measurement the $E$ field was swept between $\pm$1\,MV/m in the presence of the $H$=+30\,kOe field.
In this experimental technique, $M$ and $P$ were measured simultaneously.
This method gives a more reliable way to identify the changes in the $M$-$E$ curves.

To illustrate the agreement between the $P$ and $M$ measurements more clearly, we present the ME susceptibility $\chi/c$ in Fig.~\ref{lcpo03}(b), which is deduced from the $P$-$E$ loop by subtracting the dielectric part of the polarization $P_E=\epsilon{E}$.
Similarly to earlier research on Cr$_2$O$_3$~\cite{Iyama2013PRB}, the ME susceptibility loop ($\chi/c$-$E$) directly measures the AFM domain population, as $\chi = g \ell$.
The magnetization change in Fig.~\ref{lcpo03}(a) shows a symmetric butterly shape, while the $P$-$E$ and $\chi$-$E$ curves in Fig~\ref{lcpo03}(b) exhibit conventional ferroic hysteresis loops in shape.
The experimental observations are consistent with those illustrated in Figs.~\ref{lcpo01b}(a) and \ref{lcpo01b}(b).
The two linear parts of the $M$-$E$ butterfly loop correspond to the two AFM states.
ME susceptibilities, which correspond to the slopes of the $M$-$E$ curve on the positive and negative $E$ field sides, have the same magnitude, but the opposite sign.
The $\Delta{M}$-$E$ and $\chi/c$-$E$ hysteresis loops have exactly the same widths, and the $(EH)_C$ coercive field is defined as the condition of $\Delta{M}$=0\,$\mu_B$/f.u. or $\chi/c$=0\,ps/m.
Dependence of the $\Delta{M}$-$E$ loops on the sign of the poling fields is shown in Fig.~S2~\cite{Kocsis2020PRB2SM}, demonstrating that an arbitrary initial state can be selected and accessed by the application of appropriate $E$ and $H$ fields.

Figures~\ref{lcpo04}(a) and \ref{lcpo04}(b) show the bias $E$-field dependence of the $P$-$H$ hysteresis loops at $T$=20\,K for $\mathbf{E}\parallel{y}$, $\mathbf{H}\parallel{x}$ and $\mathbf{E}\parallel{x}$, $\mathbf{H}\parallel{y}$, respectively.
In Fig.~\ref{lcpo04}(a), the $P$-$H$ hysteresis loops have butterfly shape, similar to the $M$-$E$ measurements shown in Fig.~\ref{lcpo03}(a).
In a broad region of low $H$ fields, the $P$-$H$ has linear field dependence, while at high fields the curves switches between positive and negative ME susceptibilities.
As the bias $E$-field is increased, the coercive field $H_{\rm C}$ becomes smaller.
When $H$ field is applied along the easy-axis of LiCoPO$_4$ ($\mathbf{H}\parallel{y}$), the material enters into a paraelectric spin-flop phase~\cite{Kharchenko2010,Khrustalyov2016,Kocsis2019PRB} and $P$ disappears above 70\,kOe, as shown Fig.~\ref{lcpo04}(b).

$E$-field dependence of the coercive $H$-field for both experimental configurations ($\mathbf{E}\parallel{y}$, $\mathbf{H}\parallel{x}$ and $\mathbf{E}\parallel{x}$, $\mathbf{H}\parallel{y}$) are compared in Fig.~\ref{lcpo04}(c).
Here $H_{\rm C}$ is determined as the field where the $P$ jumps into the other branch.
As expected from the model in Sec.~\ref{Sec:Theory}, the width of $P$-$H$ hysteresis loops shows strong $E$-field dependence for both field configurations; $H_{\rm C}$ decreases as the $E$ field is increased.
In the transverse case $\mathbf{H}\parallel{x}$, $H_{\rm C}$ follows the predicted 1/$E$ dependence of Eq.~(\ref{eq:EH1}) in high $E$ fields.
This result is in accord with the previous report~\cite{Zimmermann2014}.
However, for low $E$-fields, deviation from the 1/$E$ dependence is obvious, while for the longitudinal case ($\mathbf{H}\parallel{y}$) the agreement is only qualitative.
The most important reason for the deviation from the 1/$E$ behavior is the field dependence of the $\chi$, as the ME effect is not linear any more at high fields;
The transverse component $\chi_{yx}$ decreases for high $\mathbf{H}\parallel{x}$ fields, while the $\mathbf{P}\parallel{x}$ vanishes when the $\mathbf{H}\parallel{y}$ field reaches spin-flop phase transition.
Further measurements testing the $H$-field dependence are shown in Fig.~S3.

Figures~\ref{lcpo05}(a) and \ref{lcpo05}(b) show $P$-$H$ hysteresis loops in the presence of bias field $E$=1\,MV/m at selected temperatures for $\mathbf{E}\parallel{y}$, $\mathbf{H}\parallel{x}$ and $\mathbf{E}\parallel{x}$, $\mathbf{H}\parallel{y}$, respectively.
In both experimental configurations the $P$-$H$ hysteresis show strong temperature dependence.
The coercive magnetic fields $H_{\rm C}$ are plotted in Fig.~\ref{lcpo05}(c) as a function of temperature for both cases in comparison.
Equation~(\ref{eq:EH3}) predicts temperature dependence of $H_{\rm C}\sim(T_{\rm N}-T)^{3/2}$ for the coercive $H$-field, which is adequately satisfied in the transverse case $\mathbf{H}\parallel{x}$.

    \begin{figure}[t!]
 	
    \includegraphics[width=8.0truecm]{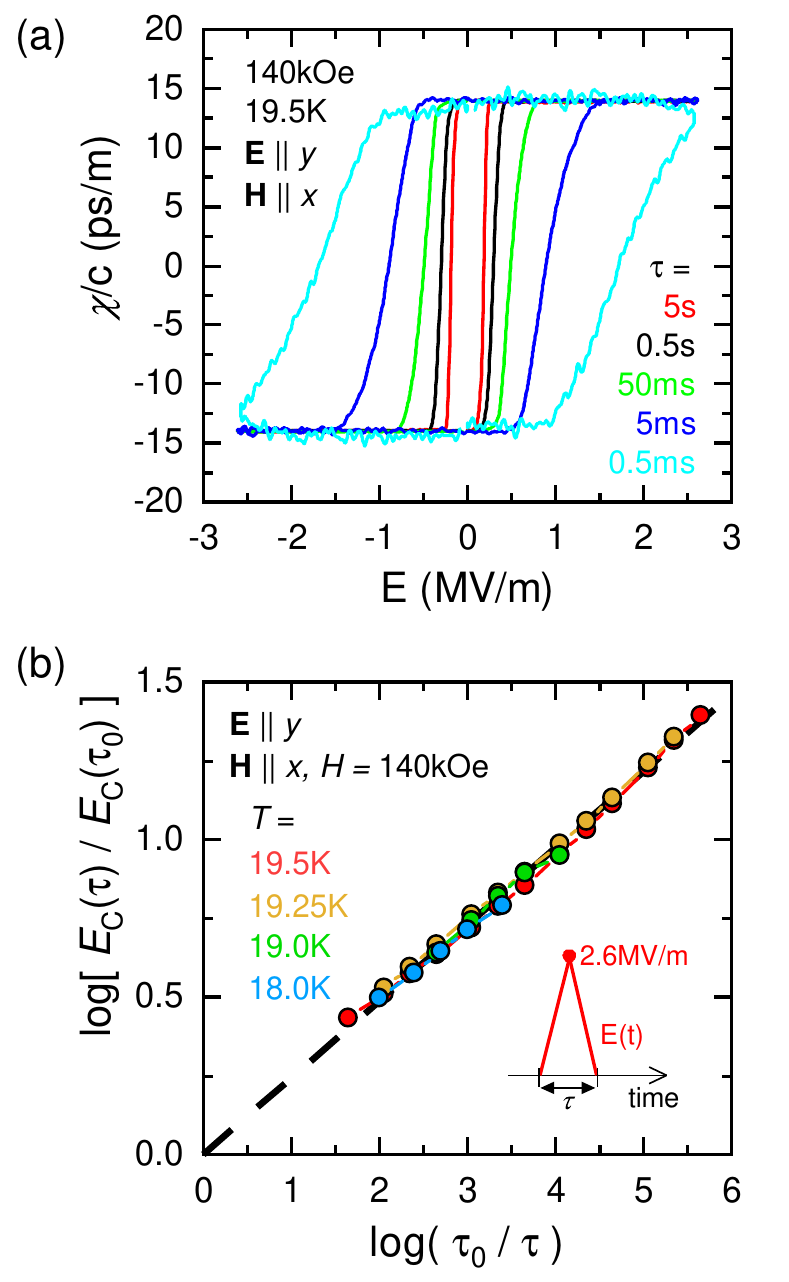}
    \caption{(Color online)
    (a) Time constant ($\tau$) dependence of the $\chi/c$-$E$ hysteresis loops at $T$=19.5\,K. The $\chi/c$ was deduced from the pulsed $E$-field PUND measurements. During the application of $E$, $H$=140\,kOe field was kept in $\mathbf{E}\parallel{y}$, $\mathbf{H}\parallel{x}$ configuration.
    (b) The pulse width dependent coercive field $E_{\rm C}(\tau)$ plotted against $1/\tau$ in a double logarithmic scale. Both $E_{\rm C}$ and $\tau$ were normalized by the corresponding values of the quasi-static measurements, $E_{\rm C}(\tau_0)$ and $\tau_0$=224\,s, respectively. Irrespective of the temperature, the $E_{\rm C}(\tau)$ - $1/\tau$ data fall on the single line in the plot, which suggests a power-law behavior. The inset of panel (b) shows the applied triangle-shaped $E$-field pulse with a width of $\tau$ and $E$=2.6\,MV/m peak amplitude.
    }
    \label{lcpo08}
    \end{figure}

\subsection*{Control over the AFM state with pulsed $E$ fields}

From the viewpoint of applications, it is crucial to investigate the dynamics of the AFM domain switching driven by rapidly changing fields.
In these experiments, triangular $E$-field pulses ($\mathbf{E}\parallel{y}$) with time duration $\tau$ as shown in the inset to Fig.~\ref{lcpo08}(b) were applied to switch the AFM states in the presence of a static field $H$=140\,kOe ($\mathbf{H}\parallel{x}$).
The $P$ of magnetic origin was measured with the PUND method.
Figure~\ref{lcpo08}(a) shows the deduced $\chi/c$-$E$ hysteresis loops for selected values of the time constant $\tau$ at $T$=19.5\,K.
Width of the $\chi/c$-$E$ hysteresis loops significantly increases for the application of shorter $E$-field pulses, \textit{i.e.} it becomes harder to switch AFM domains.
This suggests that the AFM domain walls can only propagate with a small characteristic speed, and cannot follow the rapid change of the $E$ field.
The time constant dependence of the coercive field $E_{\rm C}$ is shown at selected temperatures in a linear scale in Fig.~S4.

In Fig.~\ref{lcpo08}(b) we show the ${E}_{\rm C}(\tau) / E_{\rm C}(\tau_0)$ ratio as a function of $\tau_0/\tau$ in a log-log scale for selected temperatures.
The coercive electric field $E_{\rm C}(\tau_0)$ was measured in the quasi-static experiments, while the time constant $\tau_0$=224\,s corresponds to the time scale of the quasi-static measurements.
The duration dependent coercive field $E_{\rm C}(\tau)$ follows a power-law behavior, which is demonstrated by the linearity over a wide $\tau$-scale in the log-log plot.
Fitting of the experimental data can be conducted in the following empirical formula:
\begin{equation}
\frac{E_{\rm C}(\tau)}{E_{\rm C}(\tau_0)} =  \left(\frac{\tau}{\tau_0}\right)^{-\zeta_{\rm ME}},
\end{equation}
where the exponent is $\zeta_{\rm ME}$=0.24$\pm$0.01, which is in agreement with the Ishibashi-Orihara model of ferroelectric materials~\cite{Ishibashi1995}.
The Ishibashi-Orihara model is an extension of the Avrami–Kolmogorov model of phase transitions~\cite{Kolgomorov1937,Avrami1939,Avrami1940} to domain switching in time-dependent external fields, where the domain wall velocity $\nu$ is assumed to be solely determined by the $E$ field $\nu(t)=\nu(E(t))$, and the nucleation process is deterministic.
Although the power-law behavior is apparently independent of the temperature, this should be tested in a broader temperature range.
This $\zeta_{\rm ME}$ exponent is about $2-5$ times larger than the earlier observations in other multiferroic and ferroelectric compounds~\cite{Scott1996,Ruff2017Mats,Ruff2018APL}, and similar to triglycine-sulfate~\cite{Ishibashi1995}.
The larger $\zeta_{\rm ME}$ exponent shows the higher sensitivity of the AFM domain wall dynamics to the duration of the $E$-field pulses.
Using the Ishibashi-Orihara model, the $\chi/c$-$E$ loops are fitted with $\chi/c = \chi_0/c \cdot (1-2\exp{(-A\cdot{E}^\alpha)})$, which gives $\alpha$=4.0 for the exponent of the $E$.
The dimension $d$ of the domain wall propagation is given by the expression $d=\alpha\cdot\zeta_{\rm ME}$~\cite{Ishibashi1995}, resulting in $d$=0.96.
This means that domain walls propagate in a 1-dimensional manner by parallel translation.
This is consistent with the highly uni-axial nature of the magnetic anisotropy and the formerly observed striped antiferromagnetic domain pattern~\cite{Zimmermann2009,Zimmermann2014}.

%
%
	\section{Summary}
	
The orthorhombic LiCoPO$_4$ has a collinear antiferromagnetic (AFM) ground state with linear magnetoelectric (ME) effect, $P_\mu=\chi_{\mu\nu}H_\nu$ and $\mu_0M_\mu=\chi_{\nu\mu}E_\nu$, where $\mu,\nu = x,y$.
The ME effect of the two AFM states ($\alpha$ and $\beta$) has opposite sign, $\chi_{\mu\nu}(\alpha) = -\chi_{\mu\nu}(\beta)$, which allows for the selection or switching of the AFM state by simultaneous application of large electric ($E$) and magnetic ($H$) fields.
Once the AFM state is selected, it is preserved by an energy barrier.
Exploiting the cross-coupling between the polarization and magnetization, we have investigated the stability of the AFM phase in LiCoPO$_4$:
On the basis of a simplified phenomenological model, the product of the $E$ and $H$ coercive fields that are required to switch between the AFM states is expressed in terms of the energy barrier and ME susceptibility.
We measured the $P$ and $M$ in the simultaneous presence of high $E$ and $H$ fields.
The obtained $P$-$H$ and $M$-$E$ loops allowed us to determine the $E$- and $H$-field dependence of coercive $H$- and $E$-field, respectively, as well as the temperature dependence of the product coercive field $(EH)_{\rm C}$ for two different field configurations.
The $(EH)_{\rm C}$ field is nearly constant at a given temperature, which is in good accord with the previous report~\cite{Zimmermann2014}.
In addition, the product coercive field is found to have the temperature dependence that $(EH)_{\rm C}\sim(T_{\rm N}-T)^{3/2}$, which is predicted by the model.
Using pulsed $E$-field measurements, we also studied the dynamics of the AFM domain switching.
We found that the $(EH)_{\rm C}$ follows the power-law dependence on $\tau$ (time constant) predicted by the Ishibashi-Orihara model in the temperature range of the present experiment near $T_{\rm N}$, which implies 1-dimensional character of domain wall propagation.

\section*{Acknowledgements}
The authors are grateful for the fruitful discussions with Fumitaka Kagawa. V. Kocsis was supported by RIKEN Incentive Research Project FY2016.
Structural unit cell of the LiCoPO$_4$ crystal was illustrated using the software \texttt{VESTA}\cite{Momma2011}.


%

\newpage

\pagestyle{empty}

\newpage
\newpage

\renewcommand{\thefigure}{S\arabic{figure}}
\renewcommand{\theequation}{S\arabic{equation}}
\renewcommand{\thetable}{S\arabic{table}}
\setcounter{figure}{0}
\cleardoublepage

\begin{center}
\textbf{Supplementary Material}
\end{center}

\begin{figure}[t!]
 	
    \includegraphics[width=8.0truecm]{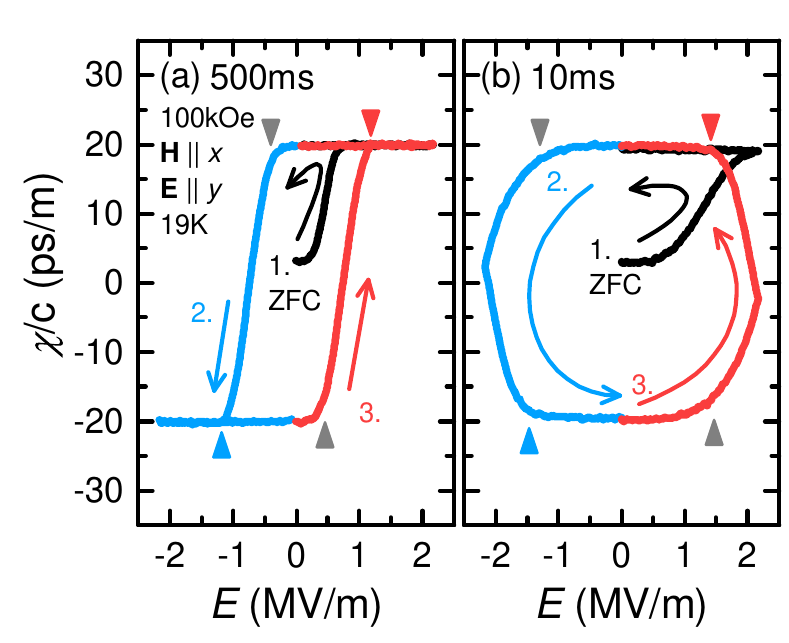}
    \caption{(Color online)
    (a-b) ME susceptibility determined from pulsed $E$-field measurements. The $P$-$E$ loops were taken in the presence of $H$=100\,kOe field at $T$=19\,K after zero field cooling (ZFC).
Starting from ZFC, the first application of the $E$ field selects a single-domain state (black curve). The $E$ and $H$ fields were applied along the $y$ and $x$ axes, respectively. For short time constant, $\tau$=10\,ms in panel (b), the hysteresis loop is incomplete as the coercive field exceeds the maximum of the applied field. However even in this case the ME-state switching was successful.}
    \label{lcpoS04}
    \end{figure}

The magnetoelectric (ME) cooling process across $T_{\rm N}$ is not the only way to form single-domain antiferromagnetic (AFM) states in LiCoPO$_4$.
Figure~\ref{lcpoS04}(a) shows a $\chi/c$-$E$ hysteresis loop measured in the presence of $H$=100\,kOe field at $T$=19\,K, with $\mathbf{E}\parallel{y}$, $\mathbf{H}\parallel{x}$ configuration, using the pulsed $E$-field PUND technique.
The experiment was started from a multi-domain AFM state, which was prepared by cooling the sample below $T_{\rm N}$ in the absence of any fields.
After the cooling, first a static $H$=140\,kOe field, and then short triangular $E$-field pulses with rise and fall time $\tau$=500\,ms were applied.
Upon the first application of the $E$-field pulse (black curve), the $\chi/c$ is irreversibly saturated and a single-domain AFM state is formed.
Further applying of the $E$-field pulses (blue and red curves), the order parameter is switched between the AFM-$\alpha$ and AFM-$\beta$ states, depending on the sign of the $E$ field.
The initial $\chi/c$-$E$ curve is saturated at $E$=0.5\,MV/m (grey triangle), at much lower field than those started from a single-domain state ($E$=1.1\,MV/m, blue and red triangle).
Moreover, even when the $E$ field is reversed, the single-domain state remains unchanged up to $\vert{E}\vert$=0.4\,MV/m magnitude of the field (grey triangles).

For shorter $E$-field pulses ($\tau$=10\,ms), the saturation and coercive fields significantly increase, as shown in Fig.~\ref{lcpoS04}(b).
In this case, the single-domain state is not reached at the maximum of the applied $E$ field.
This suggests that a single-domain AFM state is switched in two steps.
In the first step, minority-domain seeds nucleate, then in the second step switching of the AFM state is completed via the propagation of domain walls.
In the multi-domain case, the first step is less important, as there are domain walls already available, which explains the smaller saturation $E$ field for the zero-field cooled curves.

\begin{figure}[t!]
 	
    \includegraphics[width=8.0truecm]{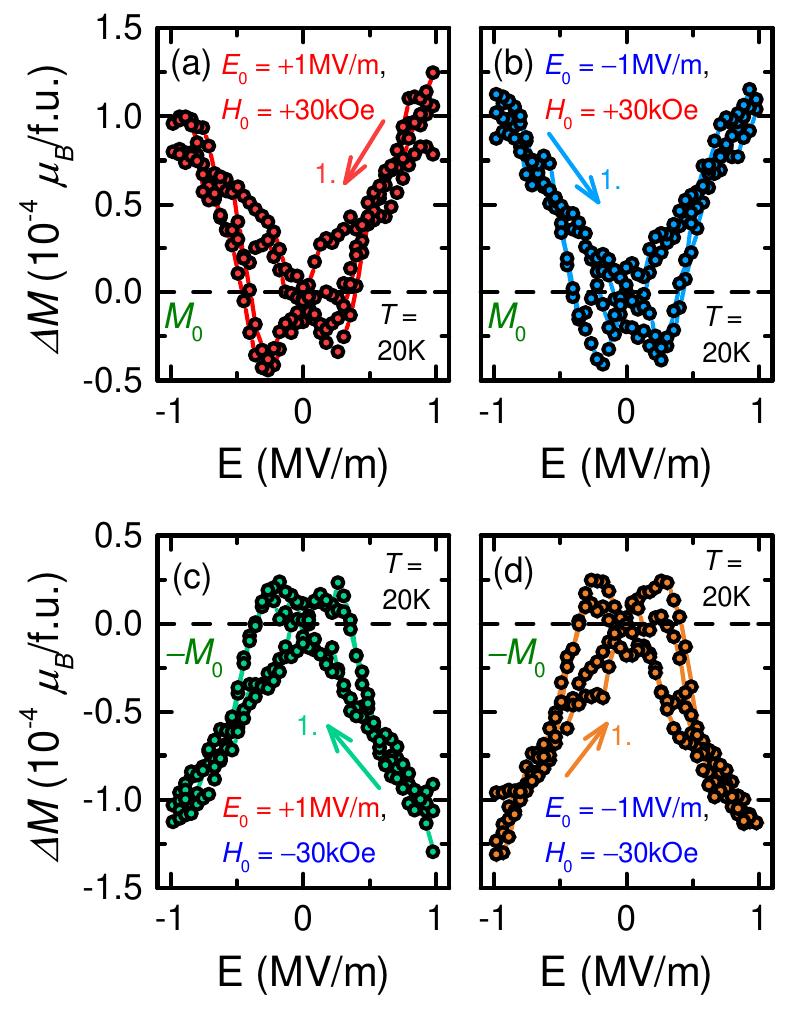}
    \caption{(Color online) Poling field dependence of the $\Delta{M}$-$E$ loops. During the poling, the samples were cooled to $T$=20\,K in the presence of $\mathbf{E}\parallel{y}$ and $\mathbf{H}\parallel{x}$ fields with the magnitudes of 1\,MV/m and 30\,kOe, respectively. Signs of the poling fields are indicated in each panel, namely ($+E_0$,$+H_0$), ($-E_0$,$+H_0$), ($+E_0$,$-H_0$), and ($-E_0$,$-H_0$) for (a) to (d), respectively. The measurements were done in the presence of the same magnetic field as applied during the poling, \textit{i.e.} $+H_0$ in case of (a,b) and $-H_0$ in case of (c,d). Starting points of the measurements are indicated by arrows.}
    \label{lcpoS02}
    \end{figure}

Figure~\ref{lcpoS02} shows the poling-field dependence of the $\Delta{M}$-$E$ loops of LiCoPO$_4$ for $\mathbf{E}\parallel{y}$ and $\mathbf{H}\parallel{x}$.
In panel (a) the measurement was started from an initial state, which was selected by cooling the sample to $T$=20\,K in the presence of ($+E_0$,$+H_0$) poling fields, with $\vert{E_0}\vert$=1\,MV/m and $\vert{H_0}\vert$=30\,kOe.
After the ME poling, two $\Delta{M}$-$E$ loops were measured in the presence of the $+H_0$ magnetic field.
The magnetization change $\Delta{M}=M(E)-M_0$ is plotted, where the base magnetization $\vert{M_0}\vert$=0.13\,$\mu_B$/f.u. is taken at $H$=+30\,kOe and $E$=0\,MV/m.
Similarly, the initial states in panels (b) to (d) were prepared by the ($-E_0$,$+H_0$), ($+E_0$,$-H_0$), and ($-E_0$,$-H_0$) combinations of the poling fields, respectively.

In panels (c,d) the measurement was performed in the presence of $-H_0$ magnetic field, and accordingly the base magnetization $M_0$ was negative.
The starting points of the measurements are indicated by arrows.
In panels (a,b) and panels (c,d), the $\Delta{M}$-$E$ curves show the same butterfly-shape hysteresis loops, as discussed in the main text, but with opposite signs.

\begin{figure}
 	
    \includegraphics[width=8.0truecm]{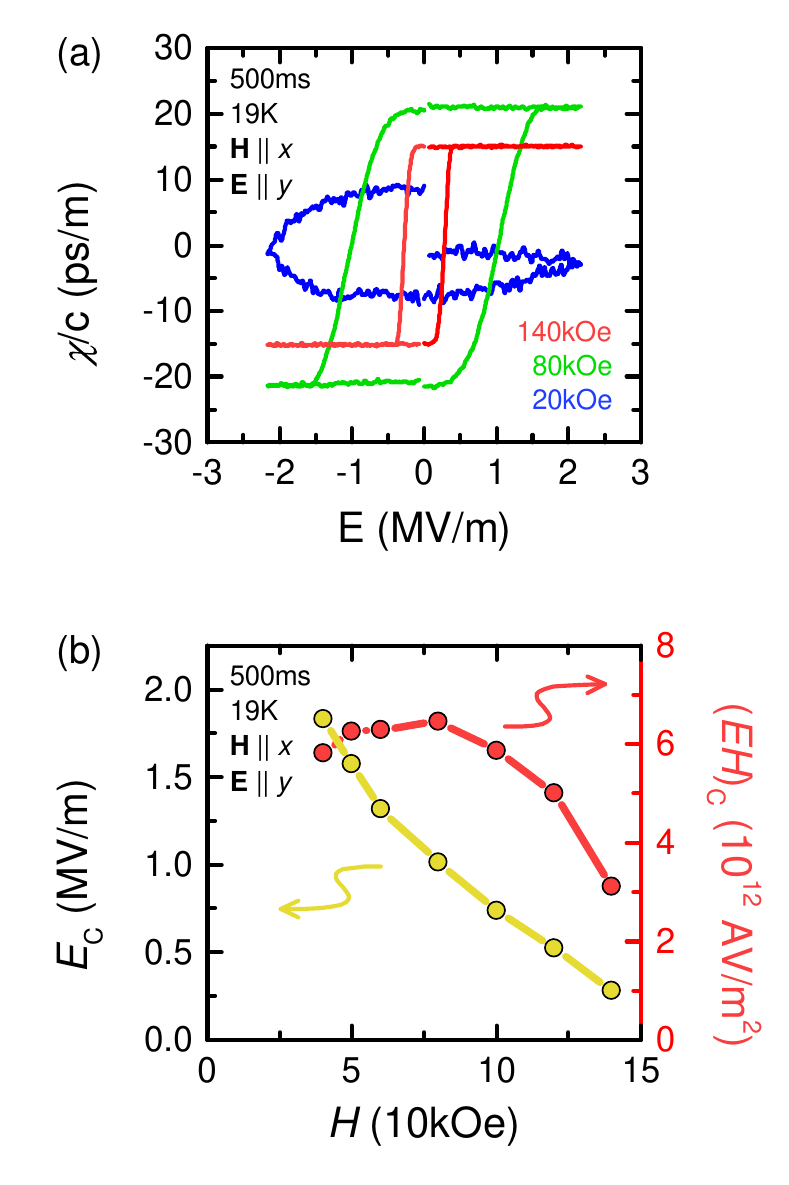}
    \caption{(Color online)
    (a) ME susceptibility deduced from the pulsed $E$-field PUND ($P$-$E$) measurements. The PUND loops were taken in the presence of different $H$ fields at $T$=19\,K, while the rise and fall time of the pulse was $\tau$=500\,ms. The $E$ and $H$ fields were applied along the $y$ and $x$ axes, respectively.
    (b) Magnetic field dependence of the coercive $E$ field needed to switch the ME states, corresponding to the data in panel (a). In the same panel we plot the $EH$ product field, which is constant only in a limited field region (50\,kOe~$<H<$~100\,kOe) far from the phase boundary of the AFM order.}
    \label{lcpoS03}
    \end{figure}

Figure~\ref{lcpoS03}(a) shows  the $H$-field dependence of the $\chi/c$-$E$ hysteresis loops measured at $T$=19\,K.
The hysteresis loops were measured with the pulsed $E$-field PUND technique, $\tau$=500\,ms, $\mathbf{E}\parallel{y}$, $\mathbf{H}\parallel{x}$.
For low $H$-fields (20\,kOe, blue curve), the hysteresis loop is open, and decreases in magnitude after one cycle, which means that the ME-state switching is incomplete.

Figure~\ref{lcpoS03}(b) shows the corresponding $H$-field dependence of the coercive electric field $E_{\rm C}$ as well as the coercive ME field $(EH)_{\rm C}$ at $T$=19\,K.
The LiCoPO$_4$ samples used in this experiment have very high coercive field with $(EH)_{\rm C}\sim$6$\cdot$\,10$^{12}$\,AV/m$^2$, much \lq\lq harder \rq\rq magnetoelectric as compared to those samples used in Ref.~\onlinecite{Zimmermann2014}.
As a result of the very high coercive $(EH)_{\rm C}$ fields, isothermal switching is only possible at temperatures just below $T_{\rm N}$.
According to the theoretical model, the $(EH)_{\rm C}$ is supposed to be independent of the $H$ field.
The decrease in the $(EH)_{\rm C}$ coercive field towards higher $H$-fields is related to the proximity to the AFM-PM first-order phase boundary, where both phases coexist. 

The experiments shown in Figs.~\ref{lcpoS04}, \ref{lcpoS02}, and \ref{lcpoS03} demonstrate controllability over the AFM order parameter via the ME effect.
Any state can be initialized and accessed either via ME cooling or via isothermal switching.

\begin{figure}[h]
 	
    \includegraphics[width=8.0truecm]{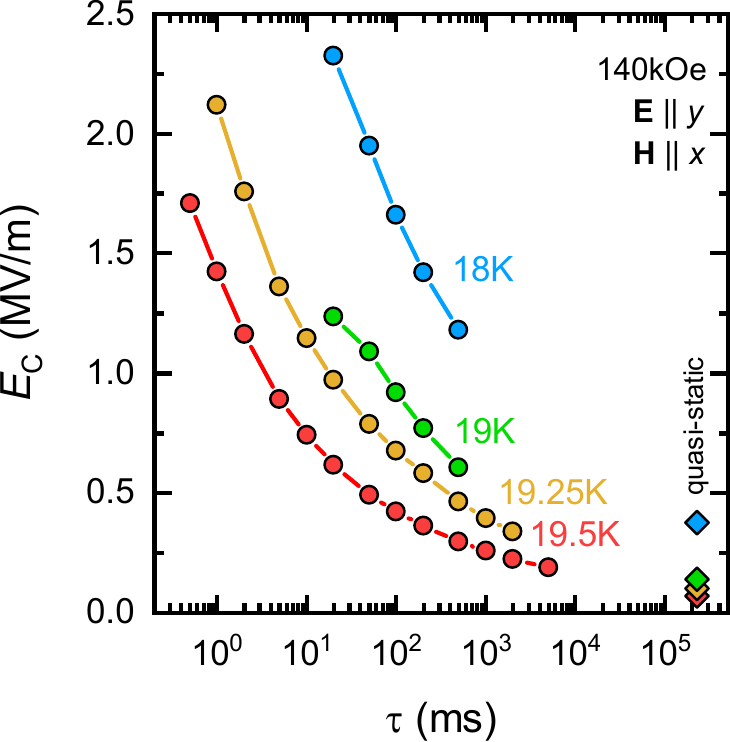}
    \caption{(Color online) Time constant ($\tau$) dependence of the coercive $E$ field at different temperatures on a semi-logarithmic scale. During the measurement, $H$=140\,kOe field was applied, $\mathbf{E}\parallel{y}$, $\mathbf{H}\parallel{x}$. Quasi-static limit of the coercive $E$ field data is indicated by bold diamonds on the right side of the plot. }
    \label{lcpoS05}
    \end{figure}

Figure~\ref{lcpoS05} shows the time constant dependence of the $E_{\rm C}$ coercive $E$-field at selected temperatures, corresponding to the data shown in Fig.~6(b) in the main text.

\newpage

\end{document}